\documentclass[usenatbib,useAMS,manuscript]{mn2e}
\usepackage{amssymb}
\usepackage{graphicx}

\title
[Testing MOND through velocity dispersion profiles]
{Testing Modified Newtonian dynamics through statistics of velocity dispersion profiles in the inner regions of elliptical galaxies} 

\author[Chae \& Gong]
{Kyu-Hyun~Chae$^{1\star}$ and In-Taek Gong$^{1}$ \\
\\
$^1$Department of Physics and Astronomy, Sejong University, 
 98 Gunja-dong, Gwangjin-Gu, Seoul 143-747, Republic of Korea\\
$^\star$chae@sejong.ac.kr}
\date{
Accepted ........;
Received .......;
in original form ......}

\pubyear{2015}

\begin{document}

\maketitle

\begin{abstract}
Modified Newtonian dynamics (MOND) proposed by Milgrom provides a 
paradigm alternative to dark matter (DM) that has been successful in fitting and
predicting the rich phenomenology of rotating disc galaxies. 
There have also been attempts to test MOND in dispersion-supported 
spheroidal early-type galaxies, but it remains unclear whether MOND can fit
 the various empirical properties of early-type galaxies for the whole ranges of
mass and radius. As a way of rigorously testing MOND in  elliptical galaxies
 we calculate the MOND-predicted velocity dispersion profiles (VDPs) 
in the inner regions of $\sim 2000$ nearly round SDSS elliptical galaxies 
under a variety of assumptions on velocity dispersion (VD) anisotropy, 
and then compare the predicted distribution of VDP slopes with the observed 
distribution in 11 ATLAS$^{\rm 3D}$ galaxies 
 selected with essentially the same criteria. 
We find that the MOND model parameterised with an interpolating function 
that works well for rotating galaxies can also reproduce
 the observed distribution of VDP slopes based only on the observed stellar 
mass distribution without DM or any other galaxy-to-galaxy varying factor. 
This is remarkable in view that Newtonian dynamics with DM requires a specific
 amount and/or profile of DM for each galaxy in order to reproduce 
the observed distribution of VDP slopes. When we analyse 
non-round galaxy samples using the MOND-based spherical Jeans equation, 
we do not find any systematic difference in the mean property of
 the VDP slope distribution compared with the nearly round sample. 
However, in line with previous studies of MOND through individual analyses
of elliptical galaxies, varying MOND interpolating function or VD anisotropy 
can lead to systematic change in the VDP slope distribution, indicating that 
a statistical analysis of VDPs can be used to constrain specific MOND models 
with an accurate measurement of VDP slopes or a prior constraint on VD 
anisotropy.
\end{abstract}


\maketitle

\section{Introduction}

Mass discrepancy (or missing mass) problems in galaxies, galaxy clusters and the
 Universe have usually been attributed to dark matter (DM) assuming that 
standard Newton-Einstein gravity is valid in those dynamical systems 
(for reviews, see, e.g., \citealt{Tri87,BHS05,San14a}). The currently 
popular Lambda cold dark matter ($\Lambda$CDM) cosmological paradigm of the 
Universe provides a successful phenomenological model of the large-scale 
structure, the cosmological microwave background radiation (CMBR) anisotropy 
power spectrum, and the expansion history of the Universe 
(e.g.\ \citealt{wmap13,plan13}). Theories of galaxy formation and evolution 
based on the $\Lambda$CDM paradigm have also been making progress with the
goal of explaining the rich phenomenology of galaxies (for a review see, e.g., 
\citealt{MBW}). 

However, in the realm (where dynamics is non-linear) of galaxies
the $\Lambda$CDM paradigm appears to face serious challenges. 
Rotation curves of disc galaxies start to deviate from Newtonian 
expectation all at the same critical acceleration scale 
$a_0 \approx 1.2 \times 10^{-10}{\rm m~s^{-2}}$ regardless of galaxy size, 
luminosity, surface brightness or any other property (\citealt{McG04,FM12}). 
The velocity at the flat part of the rotation curve is well correlated with the 
total baryonic mass of the galaxy (consistent with zero intrinsic scatter) 
spanning five orders-of-magnitude in baryonic mass from giant spiral 
galaxies to low surface brightness galaxies; this correlation is now known as
the baryonic Tully-Fisher relation (\citealt{McG05,McG11,FM12}). 
Furthermore, to every feature in an individual rotation curve there
corresponds a feature in baryonic mass density profile even in galaxies whose
dynamics should be dominated by DM if the $\Lambda$CDM paradigm is assumed
(\citealt{Sanc04,FM12}).

The apparent existence of the critical acceleration $a_0$ and the intimate 
connection between the galaxy rotation curve and 
the baryonic mass distribution seems unnatural in the $\Lambda$CDM paradigm. 
Other outstanding challenges of the $\Lambda$CDM 
paradigm in galaxies such as the number and distribution of DM 
subhaloes and the DM central cusp/core density profile are also well documented 
in the literature (\citealt{FM12,KPM12,McG14}). 

Milgrom's modified Newtonian dynamics (MOND; \citealt{Mil83}) offers an 
alternative paradigm that can by-pass the above challenges of rotating galaxies
 faced by the $\Lambda$CDM paradigm. 
The MOND paradigm posits that Newton's law of gravity
 breaks down at extremely low accelerations below the critical acceleration 
$a_0$. Then, the phenomenological existence of $a_0$, the baryonic 
Tully-Fisher relation, and so on can be unified in a single MOND law 
(see \citealt{FM12} for an extensive review). 
Whether the empirical MOND law can be ultimately explained by a fundamental
law of modified gravity or somehow by revised theories of galaxy formation under
 the $\Lambda$CDM paradigm, it is clearly worthwhile as a Kepler-like law of
galactic kinematics.
If the remarkable success of the MOND paradigm in rotating galaxies implies 
an underlying law, it must also explain the empirical properties of spheroidal 
 early-type (elliptical and lenticular) galaxies (ETGs) that are different
 from disc galaxies in dynamics and structure. 
In this respect it is quite interesting to test MOND in ETGs.  

However, testing MOND (or DM) in ETGs is far more challenging because
  they lack neutral hydrogens (the crucial dynamical tracer for disc
galaxies) and orbits of stars (and other tracers) are complex (and/or unknown). 
Nevertheless, significant efforts have been made to test MOND (and DM) in
ETGs using dynamical tracers such as stars, planetary nebulae (PNe), 
globular clusters (GCs), hot X-ray gases, satellite galaxies, and gravitational
 lensing. Despite significant efforts, it is not yet clear whether MOND can 
provide a successful phenomenological model for the dynamics of ETGs.
 Sometimes, analyses of the same galaxies often reached different conclusions. 

Based on the kinematics of PNe (out to $\sim 5 R_{\rm e}$ where $R_{\rm e}$ is 
the effective radius) of three elliptical galaxies, \cite{MS03} argued that 
MOND could explain the observed kinematics well and no DM would be needed. 
However, for the same galaxies \cite{Dek05} argued that DM would be consistent 
with the data once radially varying velocity dispersion (VD) anisotropies
 were allowed. Independent analyses of PNe kinematics of two other 
ellipticals by \cite{Tir07} and \cite{Sam10} showed that MOND could explain
the kinematics without DM.
 Based on the velocity dispersion profiles (VDPs) of stacked SDSS
 satellite galaxies around two narrow luminosity ranges of red galaxies 
\cite{KP09} and \cite{Ang08} reached opposite conclusions respectively against
and for MOND because of their different assumptions on the profiles of density 
and VD anisotropy of satellites. 
\cite{Mil12} tested MOND using the mass profiles of two X-ray bright elliptical
 galaxies deduced assuming hydrostatic equilibrium and found that MOND 
successfully reproduced the profiles over a wide acceleration range from 
$> 10 a_0$ (near the galactic centre) down to $\sim 0.1 a_0$ (over 100-200 kpc).
 Analyses of (the masses of) strong lens ETGs also produced confusing 
results, some arguing against MOND (e.g.\ \citealt{Fer12}) 
while others for MOND (e.g.\ \citealt{Chi11,San14b}).

 Based on the kinematics of GCs (out to several $R_{\rm e}$) several studies 
(\citealt{SC08,Rich08,Schu12,Sam12,Sam14}) tested MOND in a handful of 
elliptical galaxies. In particular, \cite{Sam14} carried out a systematic 
analysis of ten elliptical galaxies covering a broad range of mass. 
These studies of GC kinematics employed spherical Jeans analyses assuming 
isotropy or relatively simple anisotropies. According to these studies, as 
\cite{Sam14} emphasized, MOND alone without an additional DM component 
could not provide a successful fit for several massive slow rotators 
in their outer regions (beyond $\sim 3 R_{\rm e}$), although MOND
was generally successful for less massive fast rotators (where the slow/fast
dichotomy is in the sense by \citealt{Ems07}). 
\cite{Rich08} also advocated the need of DM under MOND in the central massive 
galaxy of the Fornax cluster. It is unclear whether these results imply 
the breakdown of MOND for a class of elliptical galaxies. 
The main limits of these studies are small sample sizes and uncertainties
in VD anisotropy. 

A rigorous test of MOND in ETGs would require (1) a statistically
 representative sample of ETGs and (2) an empirical probe of mass profile
over a wide acceleration range. Here, we propose a test of MOND in elliptical
galaxies satisfying the first requirement to a large extent and the second 
requirement only partially. We calculate MOND-based VDPs of $\sim 2000$ nearly 
round galaxies from Chae, Bernardi \& Kravtsov (2014) for 
$0.1 R_{\rm e} < r < R_{\rm e}$ using
the observed stellar mass distributions under a variety of possibilities of VD 
anisotropy. We calculate the slopes of the VDPs for $0.1 R_{\rm e}<r< R_{\rm e}$
 and compare the distribution of the slopes with the measured distribution in 
 a similarly selected sample of 11 nearly round ATLAS$^{\rm 3D}$ 
(\citealt{Cap11}) galaxies. Hence our analysis is statistical in nature. 
We describe the data and the method of analysis in section~2 and the 
results in section~3. We discuss implications of our results for MOND and
give our conclusions in section~4. 

\section{Data and method of analysis}

We consider only elliptical galaxies that are nearly round so that they 
can be analysed based on the spherical symmetry assumption with minimal error.
A statistically representative sample of nearly round galaxies can be 
drawn from the immense Sloan Digital Sky Survey (SDSS: \citealt{York00}) data 
base.\footnote{http://www.sdss.org/} 
We use a sample of SDSS galaxies defined and analysed under 
the $\Lambda$CDM paradigm by \cite{CBK14} (see also \citealt{Ber10})
 that contains photometric parameters recently investigated and measured 
by \cite{Mee13} and \cite{Mee15} based on data release (DR) 7 
data.\footnote{The \cite{Mee15} photometric 
measurements for full DR7 data can be downloaded at 
http://shalaowai.physics.upenn.edu/$\sim$ameert/fit$\underbar{~}$catalog/download/}
The sample contains 2054 elliptical galaxies 
 with a mean redshift of $\langle z\rangle\approx 0.12$ 
spanning two orders of magnitude in stellar mass that
are nearly round (projected minor-to-major axis ratio $b/a > 0.85$) and
do not possess a measurable disc. 

 Because each galaxy has no measurable disc by selection, its light 
distribution on the sky can be well-described by a S\'{e}rsic (1968) profile
\begin{equation}
\Sigma(R) \propto \exp\left[-b_{n_{\rm Ser}} 
 \left(\frac{R}{R_{\rm e}}\right)^{1/n_{\rm Ser}} \right],
\label{eq:ser}
\end{equation}
where $n_{\rm Ser}$ is referred to as S\'{e}rsic index and 
$b_{n_{\rm Ser}}= 2 n_{\rm Ser} -1/3 + 0.009876/n_{\rm Ser}$ (\citealt{PS97}).
Deprojecting equation~(\ref{eq:ser}) we obtain a volume luminosity density 
profile $\rho(r)$ and then a stellar mass density profile 
through a stellar initial mass function (IMF). (For our selected galaxies
gas is ignorable particularly because they do not possess a measurable disc.)
Each galaxy has also a measured VD $\sigma_{\rm ap}$ within the SDSS aperture of 
$R_{\rm ap} = 1.5$~arcsec radius which corresponds to a physical radius between 
$0.1 R_{\rm e}$ and $R_{\rm e}$ 
 with a mean of $\langle R_{\rm ap}/R_{\rm e}\rangle \approx 0.5$.
 Note that $\sigma_{\rm ap}$ denotes a luminosity-weighted average of 
line-of-sight velocity dispersions (LOSVDs) of all stars within the aperture.

Under the standard Newtonian dynamics the observed VD $\sigma_{\rm ap}$ may be 
reproduced by adjusting the stellar mass-to-light ($M_\star/L$) ratio or 
equivalently the stellar IMF for a flexible assumption of DM mass profile 
(even including the case of no DM) 
given the empirical intrinsic galaxy-to-galaxy scatter of IMF. This means that 
$\sigma_{\rm ap}$ (an average quantity within a region) itself may not be 
a sensitive probe of gravity or mass distribution. 
However, the VDP is a powerful probe. 
The observed VDP of an individual galaxy over a radial range can be used to 
infer the total mass profile for that range under Newtonian dynamics 
(e.g.\ \citealt{Tho07}). For a large sample of galaxies
that do not have individually measured VDPs one can, as done in \cite{CBK14},  
predict VDPs using galaxy models satisfying the available observational
constraints, and then compare the statistical distribution of VDPs with an
available empirical distribution. Such a comparison of VDP slope distributions
clearly shows that galaxies must be embedded in DM halos to match the empirical
 distribution if Newtonian dynamics is assumed to be valid at all acceleration 
scale (see \citealt{CBK14} and below).

Here we carry out a similar statistical analysis of VDP slopes under MOND. 
Our analysis is based on the spherical Jeans equation given in MOND 
(e.g.\ \citealt{Ang08,KP09}) by
\begin{equation}
\frac{d[\rho(r) \sigma_{\rm r}^2(r)]}{dr} 
+ 2 \frac{\beta(r)}{r} [\rho(r) \sigma_{\rm r}^2(r)]
= -\rho(r) g(r),
\label{eq:jeans}
\end{equation}
where $\rho(r)$ is the volume luminosity density at radius 
$r$, $\sigma_{\rm r}(r)$ is the radial stellar VD, and $\beta(r)$ is the VD 
anisotropy given by $\beta(r)=1 - \sigma_{\rm t}^2(r)/\sigma_{\rm r}^2(r)$
where $\sigma_{\rm t}(r)$ is the tangential VD in spherical coordinates.
In equation~(\ref{eq:jeans}) $g(r)$ is the MONDian (`effective' or `real') 
gravitational acceleration which differs from the Newtonian acceleration 
$g_{\rm N}(r)=G M_{\rm b}(r)/r^2$ for the baryonic (stellar here) mass
 $M_{\rm b}(r)$ within $r$. The acceleration $g$ approaches $g_{\rm N}$
 for $g/a_0 \gg 1$ but tends to $\sqrt{g_{\rm N}a_0}$ as $g/a_0 \rightarrow 0$
(hence $g(r)$ due to a point mass makes a transition from the Newtonian inverse 
square law $g(r)\propto r^{-2}$ for $g/a_0 \gg 1$ to $g(r)\propto r^{-1}$ for 
$g/a_0 \ll 1$).
The transition between the Newtonian and the MONDian regimes is parameterised by
\begin{equation}
g_{\rm N} = \mu(g/a_0) g 
\label{eq:ifmu}
\end{equation} 
where interpolating function $\mu(x)$ satisfies $\mu(x) \rightarrow 1$ for 
$x \gg 1$ and $\mu(x) \rightarrow x$ for $x \ll 1$.
The transition can also be written as the inverted relation
\begin{equation}
g = \nu(g_{\rm N}/a_0) g_{\rm N},
\label{eq:ifnu}
\end{equation} 
where $\nu(y)$ satisfies $\nu(y) \rightarrow 1$ for $y \gg 1$ and 
$\nu(y) \rightarrow y^{-1/2}$ for $y \ll 1$.

We consider a class of interpolating functions given by 
\begin{equation}
 \mu_n(x)=\frac{x}{(1+x^n)^{1/n}}
\label{eq:ifmun}
\end{equation}
with the corresponding inverted function (\citealt{MS08})
\begin{equation}
 \nu_n(y)=\left[\frac{1+(1+4 y^{-n})^{1/2}}{2} \right]^{1/n},
\label{eq:ifnun}
\end{equation}
where the case $n=2$ is traditionally known as the `standard' function while the
`simple' case $n=1$ introduced by \cite{FB05} has turned out performing well in
 various recent studies (e.g.\ \citealt{SN07,Ang08,Mil12}).  
We also consider the interpolating function implied by Bekenstein's relativistic
 theory of modified gravity given by (\citealt{Bek04,ZF06})
\begin{equation}
 \mu_{\rm Bek}(x)=\frac{\sqrt{1+4x}-1}{\sqrt{1+4x}+1}
\label{eq:ifmubek}
\end{equation}
with the corresponding inverted function
\begin{equation}
 \nu_{\rm Bek}(y)=1 + y^{-1/2}.
\label{eq:ifnubek}
\end{equation}

For MONDian dynamical analyses of a galaxy any external field in which the
galaxy is embedded should, in principle, be taken into account. A MONDian
external field effect (EFE; \citealt{Mil83,Fam07,Rich11}) may particularly 
matter for those in the central regions of galaxy clusters and those having
close neighbors. Most galaxies in our SDSS sample are not in cluster centres 
and have light distributions ``uncontaminated'' by neighbors. 
Also, an evaluation of EFEs for elliptical 
galaxies using the Virgo and Coma clusters by \cite{Rich11} indicates that
the expected effects are small. In this work we do not consider including 
EFEs for our selected galaxies.

For a given baryonic mass profile $M_{\rm b}(r)$ with the corresponding 
luminosity density $\rho(r)$ the solution of equation~(\ref{eq:jeans}) 
for the radial stellar VD $\sigma_{\rm r}(r)$ can be
given following appendix~B of \cite{Chae12} as
\begin{equation}
\sigma^2_{\rm r}(r)=G
   \int_r^\infty \frac{\omega(t)}{\omega(r)}  
   \frac{\rho(t)}{\rho(r)} \frac{M_{\rm b}(t)}{t^2} 
       \nu\left(\frac{G}{a_0}\frac{M_{\rm b}(t)}{t^2}\right) dt, 
\label{eq:sigr}
\end{equation}
where $\omega(r)=\exp\left[\int^r(2\beta(r')/r')dr'\right]$ for an anisotropy
$\beta(r')$ and $\nu(y)$ is an inverted interpolating function.
The LOSVD of stars at projected radius $R$ on the sky is then 
given by (e.g.\ \citealt{BM82})
\begin{equation}
\sigma_{\rm los}^2(R)=\frac{1}{\Sigma(R)} \int_{R^2}^{\infty}
\rho(r) \sigma_{\rm r}^2(r) \left[ 1 - \frac{R^2}{r^2} \beta(r) \right]
 \frac{dr^2}{\sqrt{r^2-R^2}},
\label{eq:losvd}
\end{equation}
where $\Sigma(R)$ is given by equation~(\ref{eq:ser}).
 The luminosity weighted LOSVD within projected radius $R$ is given as
\begin{equation}
\sigma(R) \equiv \langle \sigma_{\rm los} \rangle (R)  = 
 \frac{\int_0^R \Sigma(R') \sigma_{\rm los}(R') R' dR'}
{\int_0^R \Sigma(R') R' dR'},
\label{eq:lwvd}
\end{equation}
and thus the aperture VD is $\sigma_{\rm ap}=\sigma(R=R_{\rm ap})$.

 Three factors can be involved in the connection between
the observationally derived luminosity density $\rho(r)$ and 
the observed VD $\sigma_{\rm ap}$: the stellar mass-to-light ratio $M_\star/L$, 
the VD anisotropy profile $\beta(r)$ and 
the (inverted) interpolating function $\nu(g_{\rm N}(r)/a_0)$.  
\cite{Ber10} provide stellar masses (or stellar mass-to-light ratios) of 
SDSS ETGs based on the \cite{Chab03} IMF. Following the literature (e.g.\ 
\citealt{CBK14}) we define a stellar mass-to-light ratio or IMF mismatch 
parameter 
\begin{equation}
\delta_M\equiv\log_{10}(M_\star/M^{\rm Ch}_\star) = 
\log_{10}\left[\frac{(M_\star/L)}{(M_\star/L)^{\rm Ch}}\right]
\label{eq:dm}
\end{equation}
between the unknown stellar mass $M_\star$ (or $M_\star/L$) and 
the fiducial stellar mass $M^{\rm Ch}_\star$ [or $(M_\star/L)^{\rm Ch}$] 
derived by \cite{Ber10} based on the \cite{Chab03} IMF. 
All the uncertainties related to photometric measurement, mass-to-light ratio, 
and/or IMF are then absorbed into this single parameter $\delta_M$.
Since our goal is to test MOND i.e.\ its interpolating functions, 
we need empirical inputs or assumptions 
for $\delta_M$  and $\beta(r)$.\footnote{If a VDP is measured for a radial range
 of a galaxy, then one can in principle simultaneously solve for $\delta_M$
and $\beta(r)$ for that range for the measured $\rho(r)$ and $\sigma(R)$.}
 Here we take an approach similar to \cite{CBK14}.
The empirical input for $\delta_M$ is provided by the literature results on 
VD-dependent stellar IMF (e.g. \citealt{Cap13b,CvD12,Tor13}; see below).
 For $\beta(r)$ we consider both constant $\beta$ 
and double Osipkov-Merritt-type (\citealt{Chae12,Chae14})
$\beta_{\rm dOM}(r)=\beta_1/(1+r_1^2/r^2)+\beta_2/(1+r_2^2/r^2)$, 
 where $\beta_1 + \beta_2=\beta(r\rightarrow\infty)\equiv \beta_\infty$.
 For constant $\beta$ each value is drawn from a Gaussian distribution. 
For $\beta_{\rm dOM}(r)$ the radial mean for $0\le r \le R_{\rm e}$,
referred to as $\beta_{\rm e}$,\footnote{The relation between 
$(\beta_{\rm e},~\beta_{\infty})$ and $(\beta_1,~\beta_2)$ can be found in
appendix~B of \cite{CBK14}} 
and the value at infinity ($\beta_\infty$) 
are drawn from a Gaussian distribution and $r_1$ and $r_2$ are randomly
 assigned satisfying $0 < r_1/R_{\rm e} < r_2/R_{\rm e}< 1$. The adopted function
 $\beta_{\rm dOM}(r)$ is intended to encompass (not exhaustively) possible 
behaviours of anisotropy. 

For input distributions of $\delta_M$ and $\beta(r)$ (constant or 
double Osipkov-Merritt-type) we search randomly for a pair reproducing 
$\sigma_{\rm ap}$ within a typical measurement error of $0.04$~dex for the 
observationally derived $\rho(r)$ of each galaxy. For most galaxies 
a pair is easily found. For a few per cent of galaxies a successful pair
could not be found after a significant number of trials.
For each successful model we calculate a VDP $\sigma(R)$ 
(equation~\ref{eq:lwvd}) as a function of radius $R$ 
on the plane of the sky. We use a power-law approximation 
\begin{equation}
\frac{\sigma(R)}{\sigma_{\rm e2}} = \left(\frac{R}{R_{\rm e}/2}\right)^\eta
\label{eq:eta}
\end{equation}
for $0.1R_{\rm e}< R < R_{\rm e}$, where 
$\sigma_{\rm e2}\equiv \sigma(R=R_{\rm e}/2)$ and the value of $\eta$ for each 
model is determined through a least-square fit.
The statistical distribution of $\eta$ for the galaxy models is then compared 
with an empirical distribution.

\begin{figure} 
\begin{center}
\setlength{\unitlength}{1cm}
\begin{picture}(9,6)(0,0)
\put(-0.7,7.7){\includegraphics{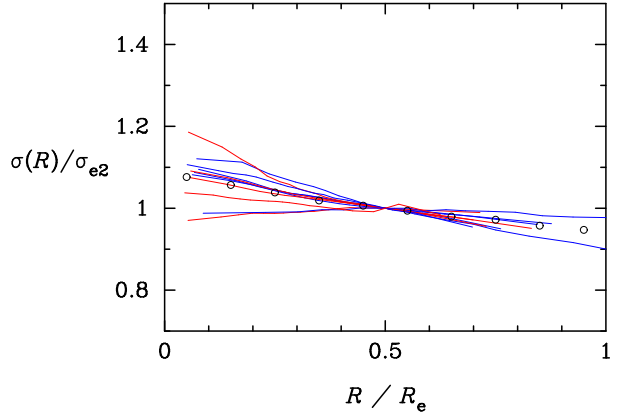}}
\end{picture}
\caption{
 Profiles of luminosity-weighted LOSVD $\sigma(R)$ (equation~11) 
for 11 nearly round ATLAS$^{\rm 3D}$ galaxies for which LOSVDs have been 
measured over 300 bins in the optical region (see the text).
 Red and blue profiles represent respectively slow and fast rotators.
The abscissa is radius $R$ on the sky normalized by the effective radius 
$R_{\rm e}$ while the ordinate is normalized by 
$\sigma_{\rm e2} \equiv \sigma(R=R_{\rm e}/2)$. Small circles represent
the mean of the displayed profiles.
}
\label{VDPA3D}
\end{center}
\end{figure}
 Published values of $\eta$ for dozens of
 ETGs have a mean of $\langle\eta\rangle = -0.06\pm 0.01$ with a
galaxy-to-galaxy intrinsic scatter of $\sigma_\eta = 0.03-0.04$ 
(e.g.\ \citealt{CBK14,Cap06}). For a more reliable comparison with our
nearly round galaxies we select galaxies from the 
ATLAS$^{\rm 3D}$ (\citealt{Cap11})
public data\footnote{http://www-astro.physics.ox.ac.uk/atlas3d/} 
using the same criteria. Out of 260 ATLAS$^{\rm 3D}$ ETGs 
 33 galaxies satisfy projected $b/a>0.85$ in both 
photometric (\citealt{Cap13a}) and kinematic (\citealt{Ems11}) distributions
within both $R_{\rm e}/2$ and $R_{\rm e}$. 
Seventeen of them are elliptical and out of these
 fifteen have measured values of LOSVD over 300  bins (pixels) on the 
plane of the optical region allowing reliable measurement of $\eta$. Finally,
excluding galaxies having the mean of bulge and total S\'{e}rsic indices
 $n_{\rm Ser}<2.5$ as for the SDSS ETG sample (\citealt{CBK14}) 
we are left with 11 galaxies:  they are  NGC 
 3193, 3379, 4168, 4278, 4283, 4374, 4458, 4552, 5173, 5638, and 5846.
For each galaxy about 20 concentric rings are defined within $R_{\rm e}$. 
LOSVDs of the bins in each ring are averaged and weighted with
the surface brightness (\citealt{Sco13}) of the ring and then these 
luminosity-weighted average LOSVDs give a $\sigma(R)$ profile. 
The derived $\sigma(R)$ profiles for the
 11 galaxies are displayed in Fig.~\ref{VDPA3D}. 
For these galaxies the least-square
fit power-law slopes for $0.1R_{\rm e}< R <R_{\rm e}$ have a mean of 
{\bf $\langle\eta\rangle = -0.057\pm 0.011$} with a standard deviation of 
{\bf $0.037\pm 0.007$} (where the quoted errors have been estimated from 
bootstrap resampling) consistent with published results for ETGs of any 
ellipticity (\citealt{CBK14,Cap06}).
\begin{figure} 
\begin{center}
\setlength{\unitlength}{1cm}
\begin{picture}(9,5)(0,0)
\put(-0.3,6.5){\includegraphics{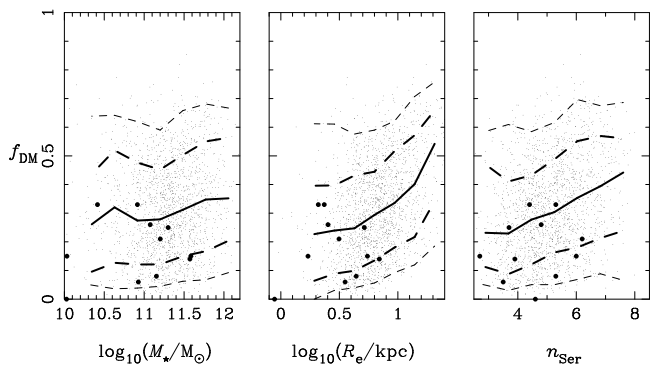}}
\end{picture}
\caption{
Dark matter fraction within the sphere of radius equal to $R_{\rm e}$ 
($f_{\rm DM}$) in  SDSS and  
ATLAS$^{\rm 3D}$ nearly round galaxies as a function of stellar mass ($M_\star$),
effective radius ($R_{\rm e}$) and S\'{e}rsic index ($n_{\rm Ser}$). Large dots
represent 11 ATLAS$^{\rm 3D}$ galaxies. SDSS galaxies are represented by small 
dots with thick solid curves representing the median values, thick and thin 
dashed curves the 68\% and 95\% limits respectively. See the text for details.
}
\label{Gdist}
\end{center}
\end{figure}

\cite{Ems11} classify ATLAS$^{\rm 3D}$ ETGs by the 
degree of large-scale rotation using the criterion defined by \cite{Ems07}. 
As noted in section~1, some previous tests (based mostly on GC kinematics) of
MOND in elliptical galaxies obtained contrasting results for fast and slow 
rotators [MOND was not successful for the outer part ($\ga 2$--$3 R_{\rm e}$) 
kinematics of slow rotators in several cases; see \cite{Sam14} and references 
therein]. For SDSS ETGs this kinematic information is missing. The majority 
($86\pm 2$ per cent) of ATLAS$^{\rm 3D}$ ETGs are classified as fast rotators.
However, for the above 11 ATLAS$^{\rm 3D}$ ellipticals selected 
using essentially the same criteria as for the SDSS sample, 
 about one half (6 out of 11) are fast rotators. 
These two kinematic classes of galaxies have mean VDP slopes 
$\langle\eta\rangle=-0.067\pm 0.014$ (fast) versus $-0.046\pm 0.017$ (slow) 
where the discrepancy is not statistically significant based on the small 
samples. We caution, however, that the low precision of current samples does 
not allow a sensitive test. 

Fig.~\ref{Gdist} shows the distributions of dark matter fraction within the 
effective radius ($f_{\rm DM}$) for the SDSS and the ATLAS$^{\rm 3D}$ nearly round
 galaxies as a function of stellar mass ($M_\star$), effective radius 
($R_{\rm e}$) and S\'{e}rsic index ($n_{\rm Ser}$), where 
SDSS parameter values are taken from the \cite{CBK14} modelling result for
their fiducial inputs. There are significant differences between 
SDSS and ATLAS$^{\rm 3D}$ samples. SDSS galaxies have larger $f_{\rm DM}$ overall 
(mean $\langle f_{\rm DM} \rangle \approx 0.32$ versus $0.18$) and are 
biased towards greater stellar masses 
(mean $\log_{10}(M_\star/{\rm M}_\odot)\approx 11.3$ versus $10.8$)
 and (more noticeably) larger sizes
(mean $\log_{10}(R_{\rm e}/{\rm kpc})\approx 0.82$ versus $0.48$). 
The range of modelling results considered by \cite{CBK14} taking into
account various systematic effects give $0.27 \la f_{\rm DM} \la 0.38$.
The second panel of Fig.~\ref{Gdist} indicates that the difference in 
$\langle f_{\rm DM} \rangle$ can be in large part attributed to the difference
in galaxy sizes.  Indeed, at $\log_{10}(R_{\rm e}/{\rm kpc})=0.5$
 SDSS galaxies have $f_{\rm DM}\approx 0.25$ (with a systematic error of
$\sim 0.05$ due to input variations) getting closer to 
$0.18$ of ATLAS$^{\rm 3D}$ galaxies. 
\begin{figure} 
\begin{center}
\setlength{\unitlength}{1cm}
\begin{picture}(9,6)(0,0)
\put(-0.7,6.5){\includegraphics{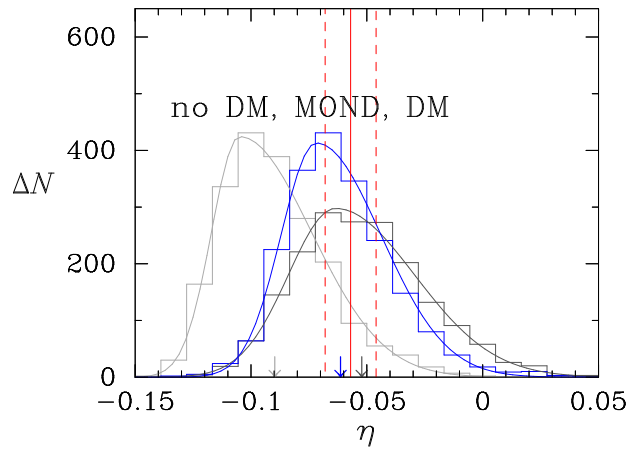}}
\end{picture}
\caption{
Predicted distributions of VDP slope $\eta$ (equation~13) for $\sim 2000$
 nearly round SDSS elliptical galaxies. Three cases are compared: Newtonian 
cases without and with DM presented in Chae et al.\ (2014) and a MOND case 
calculated here for $(n,\langle\beta\rangle)=(1.5,0)$ and shown in 
Fig.~\ref{VDPconst}. 
Vertical red solid and dashed lines indicate the measured mean and its
error for 11 ATLAS$^{\rm 3D}$ nearly round galaxies shown in Fig.~\ref{VDPA3D}. 
The arrows indicate the mean values of the distributions.
}
\label{VDPDM}
\end{center}
\end{figure}

The difference in $R_{\rm e}$ between the SDSS and the 
ATLAS$^{\rm 3D}$ samples may be due to various factors including 
different parent populations, measurement errors and different methods of
modelling light profiles. The ATLAS$^{\rm 3D}$ parent sample is a volume limited
sample in the local Universe within $D < 42$~Mpc (almost) complete  
for $M_\star \ga 6 \times 10^9 {\rm M}_\odot$ (\citealt{Cap11}). 
The SDSS parent sample is magnitude-limited to cosmological distances 
(up to redshift $\sim 0.2$ with a median redshift of $\sim 0.1$) 
and thus completeness becomes less secure  for
less luminous galaxies and at larger distances. It is also likely that 
measurement/modelling errors are at work as can be seen, e.g., from the 
systematic difference between RC3 and 2MASS photometric data for the same 
ATLAS$^{\rm 3D}$ galaxies (\citealt{Cap11}) and dependence on modelling 
details for SDSS galaxies (\citealt{Mee13}).  
\begin{figure*} 
\begin{center}
\setlength{\unitlength}{1cm}
\begin{picture}(12,11)(0,0)
\put(-1.8,12.5){\includegraphics{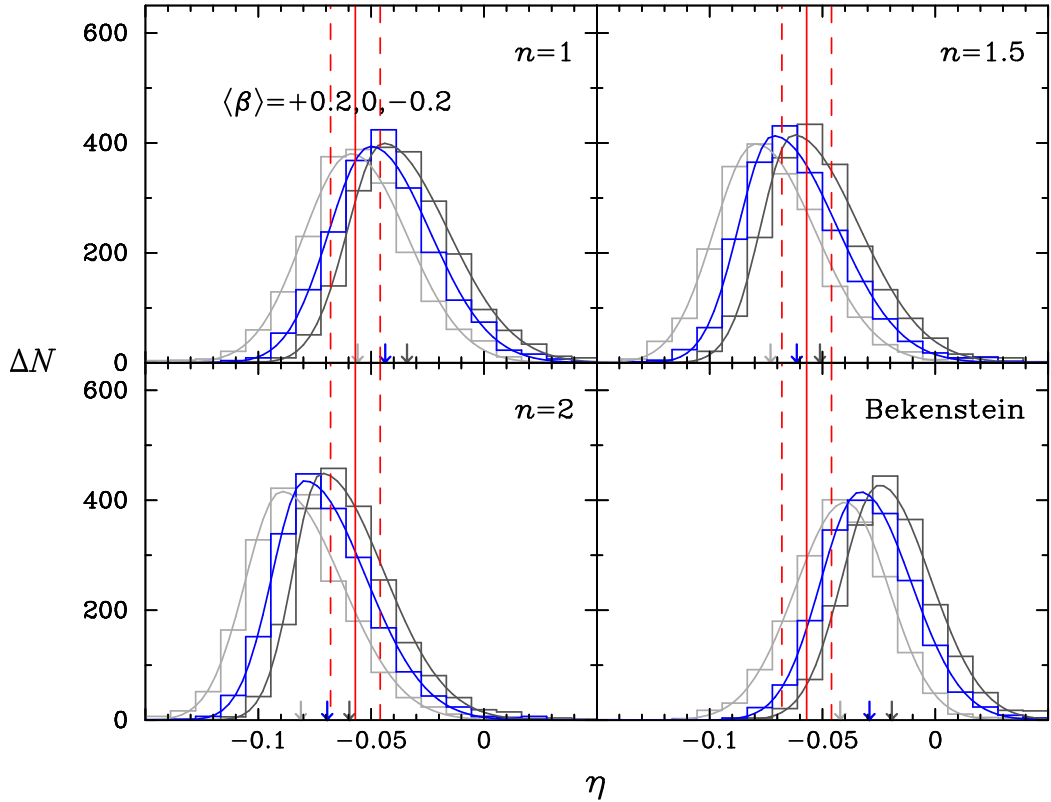}}
\end{picture}
\caption{
Predicted distribution of VDP slope $\eta$ (equation~13) for the SDSS galaxy 
sample depending on  MOND interpolation function and VD anisotropy 
mean value $\langle\beta\rangle$ assuming anisotropy is constant 
and has a galaxy-to-galaxy scatter of $0.2$.  
We consider a MOND interpolation function of the form given by equation~(5)  
with index $n=1$, $1.5$ and $2$ as well as that given by equation~(7) 
taken from Bekenstein's (2004) relativistic theory of modified gravity. 
For each MOND model we consider three cases of $\langle\beta\rangle=+0.2$ 
(light gray), $\langle\beta\rangle=0$ (blue) and $\langle\beta\rangle=-0.2$
(dark gray). Red lines are the same as in Fig.~\ref{VDPDM}.
The arrows indicate the mean values of the distributions.
}
\label{VDPconst}
\end{center}
\end{figure*}

We assume that the statistical distribution of $\eta$ is not significantly
affected by such sample difference as that between SDSS and ATLAS$^{\rm 3D}$ 
samples and use the measured distribution of $\eta$ for the ATLAS$^{\rm 3D}$ 
galaxies to interpret modelling results for the SDSS galaxies.
 We also make no distinction between fast and slow rotators and note that
our selected ATLAS$^{\rm 3D}$ galaxies are nearly evenly divided 
(Fig.~\ref{VDPA3D}).

\section{Results}

The output distribution of $\eta$ for the SDSS galaxy sample can depend on the
input stellar IMF (distribution), parameterised by $\delta_M$ 
(equation~\ref{eq:dm}), and VD anisotropy $\beta(r)$ for a given 
MOND model (interpolating function).   
Our standard choice for the VD-dependent stellar IMF is that preferred by 
ATLAS$^{\rm 3D}$ (\citealt{Cap13b}), 
 which can be re-expressed in our adaptation as 
$\delta_M=0.11+0.35\log_{10}(2^{-0.06}\sigma_{\rm e2}/130~{\rm km}~{\rm s}^{-1})$.
Unless specified otherwise, all presented results are based on the input of 
this ATLAS$^{\rm 3D}$ IMF distribution.  However, as demonstrated below,
it turns out that the output $\eta$ distribution is only weakly dependent on 
the input $\delta_M$ distribution.

We first compare the Newtonian cases without and with DM presented in 
\cite{CBK14} with a MOND case. The results for the three cases are shown in
Fig.~\ref{VDPDM}. The Newtonian case with DM reproduced here is the 
result based on the fiducial inputs by \cite{CBK14}. Newtonian dynamics 
clearly requires DM in a statistical sense (see also Fig.~\ref{Gdist})
and can match the observed VDP distribution with adjustment
 of various inputs (\citealt{CBK14}). However, it is also evident that MOND
can reproduce well the VDP distribution.

We present in turn MOND results for constant anisotropy (section~3.1) and for 
varying anisotropy of the form $\beta_{\rm dOM}(r)$ (section~3.2).

\subsection{Constant anisotropy}

\begin{figure} 
\begin{center}
\setlength{\unitlength}{1cm}
\begin{picture}(8,13)(0,0)
\put(-1.5,-0.6){\includegraphics{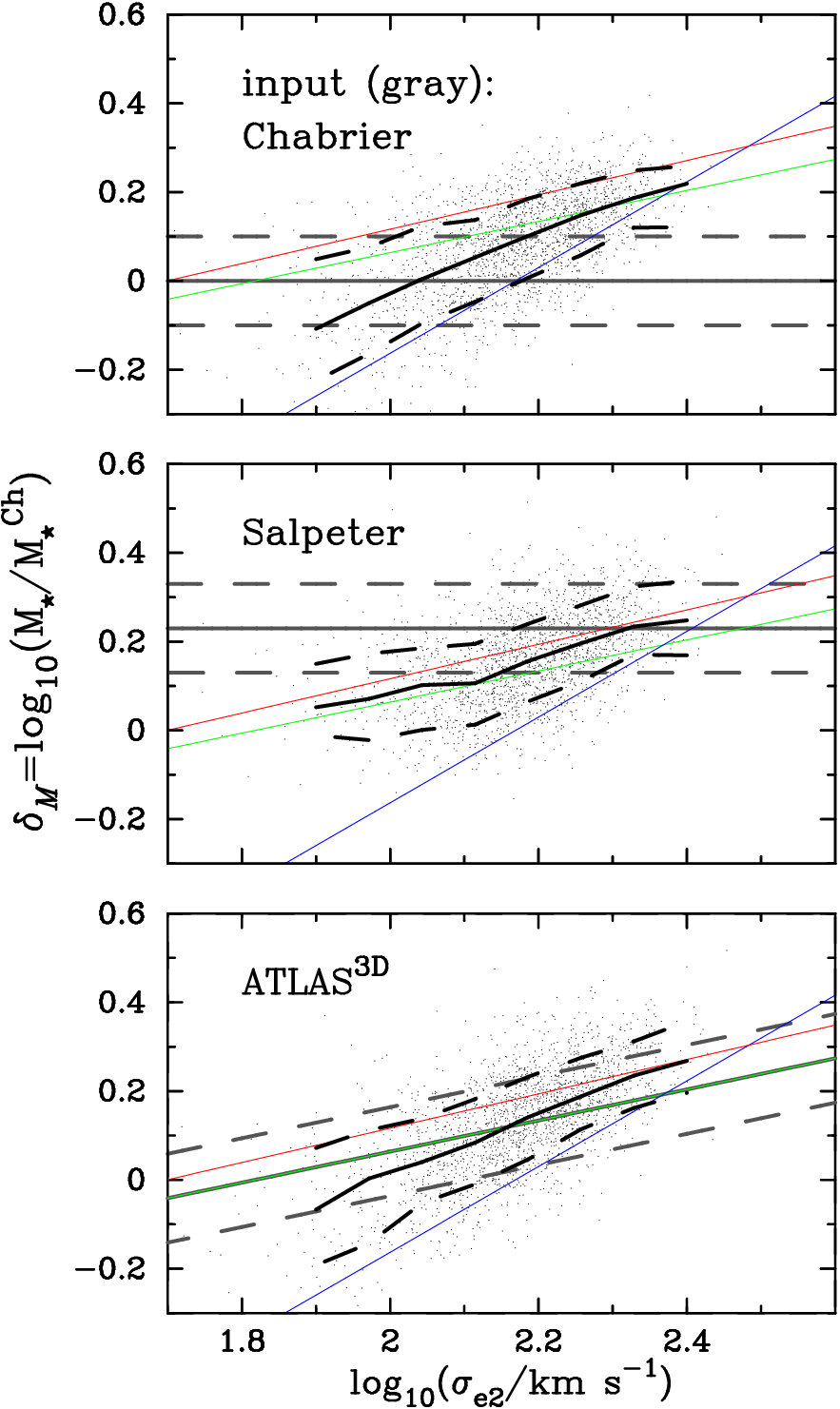}}
\end{picture}
\caption{
Input and output distributions of stellar IMF mismatch parameter 
$\delta_M$ (equation~12). 
Three input IMFs are represented by gray lines with an intrinsic scatter of 
0.1~dex (represented by dashed lines): 
Chabrier (2003), Salpeter (1955) and ATLAS$^{\rm 3D}$ (Cappellari et al.\ 2013b).
Data points are the output models and the black lines indicate the median 
relations (with dashed lines indicating the 68 percent limits). 
Thin lines represent recently measured IMF behaviours: 
red -- Conroy \& van Dokkum (2012); green -- Cappellari et al.\ (2013b);
 blue -- Tortora et al.\ (2013).
}
\label{IMF}
\end{center}
\end{figure}
  Anisotropy values are drawn from a Gaussian distribution with 
 a mean of $\langle\beta\rangle$ and a standard deviation of $0.2$, 
which is motivated from studies of
 nearby early-type galaxies (e.g.\ \citealt{Ger01,Cap07,Tho07}). 
  The predicted distribution of $\eta$ for the SDSS galaxy sample then 
depends on the input mean anisotropy $\langle\beta\rangle$
 and MOND interpolating function.
We consider $\langle\beta\rangle=+0.2$ (light gray), $0$ (blue), 
and $-0.2$ (dark gray) for each of 
$n=1$, $1.5$, and $2$ in the function given by 
equation~(\ref{eq:ifmun}) as well as for the function  given by 
equation~(\ref{eq:ifmubek}). The results are displayed in Fig.~\ref{VDPconst}.
For $n=1.5$ the predicted mean of $\eta$ can match easily the measured value.
The cases $n=1$ and $n=2$ can also match the measured 
distribution but not for all the cases of $\langle\beta\rangle$.
The case $n=1$ prefers  $\langle\beta\rangle=+0.2$ while $n=2$ prefers
 $\langle\beta\rangle=-0.2$. These results imply that a number of degenerate 
sets of $(n,~\langle\beta\rangle)$ can be consistent with the measured $\eta$ 
value. However, Bekenstein's model is disfavoured
for the assumed cases of constant anisotropies.
 
In our approach it turns out that the output distribution of $\eta$ is only  
weakly dependent on the input IMF.
In the random process of finding $(\delta_M, \beta)$ where
$\delta_M$ (equation~\ref{eq:dm}) is essentially a representation 
of the stellar mass-to-light ratio depending on the IMF,  
the posterior distribution of $\delta_M$ generally deviates from the prior 
input distribution. Fig.~\ref{IMF} shows the distributions of $\delta_M$ for
 three cases of input IMF: two VD-independent cases of \cite{Chab03} and 
 \cite{Sal55} and one VD-dependent case of ATLAS$^{\rm 3D}$ (\citealt{Cap13b}), 
with an intrinsic scatter of $0.1$~dex (\citealt{Cap13b}) imposed in all cases.
 Irrespective of the input IMF the posterior 
mean of $\delta_M$ increases with VD and its functional behaviour is well within
 the behaviours of recently inferred IMFs (\citealt{CvD12,Cap13b,Tor13}). 

 For the cases of constant anisotropy
 the predicted width (containing 68 percent of galaxies) and standard 
deviation of $\eta$ ($\sim 0.23$) are smaller than the 
measured standard deviation of $\sigma_\eta\approx 0.037\pm 0.007$. 
This discrepancy is largely removed if varying anisotropies are used 
as described in section~3.2.

\subsection{Varying anisotropy}

\begin{figure} 
\begin{center}
\setlength{\unitlength}{1cm}
\begin{picture}(9,6)(0,0)
\put(-0.9,6.5){\includegraphics{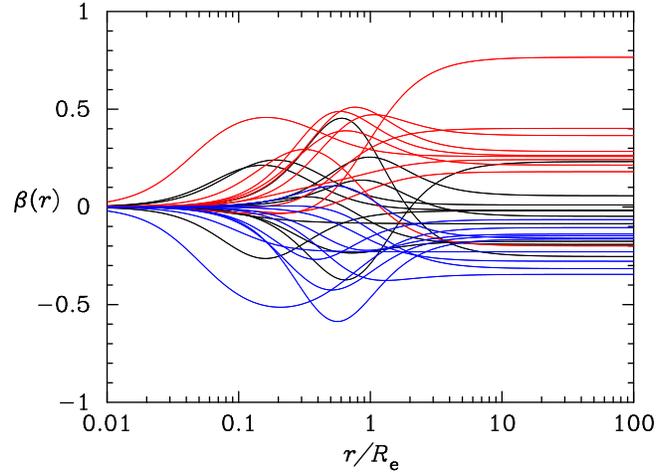}}
\end{picture}
\caption{
Examples of varying anisotropy  
$\beta_{\rm dOM}(r)=\beta_1/(1+r_1^2/r^2)+\beta_2/(1+r_2^2/r^2)$ for mean 
$\langle\beta_{\rm e}\rangle=\langle\beta_{\infty}\rangle=+0.2$ (red), $0$ (black)
 and $-0.2$ (blue) and $0 < r_1/R_{\rm e} < r_2/R_{\rm e}< 1$.
}
\label{betr}
\end{center}
\end{figure}
For $\beta_{\rm dOM}(r)=\beta_1/(1+r_1^2/r^2)+\beta_2/(1+r_2^2/r^2)$ the 
radial mean for $0\le r\le R_{\rm e}$ $\beta_{\rm e}$ and the anisotropy 
at infinity $\beta_{\infty}=\beta_1+\beta_2$ are assumed to be drawn from a 
Gaussian distribution with a standard deviation of $0.2$. 
 Because $\beta_{\rm e}$ can differ from $\beta_{\infty}$ for a galaxy 
(although the means for the entire sample are assumed to be the same, i.e.\ 
$\langle\beta_{\rm e}\rangle=\langle\beta_{\infty}\rangle$)
and $0 < r_1/R_{\rm e} < r_2/R_{\rm e}< 1$ the adopted function allows wild 
variation within $R_{\rm e}$. This does not necessarily mean that 
$\beta_{\rm dOM}(r)$ represents or fully encompasses real behaviours of 
anisotropy. We just use $\beta_{\rm dOM}(r)$ to mimic some uncertainties in 
varying anisotropies. Examples (taken from modelling results) of 
$\beta_{\rm dOM}(r)$ for mean 
$\langle\beta_{\rm e}\rangle=\langle\beta_{\infty}\rangle=+0.2$, $0$ and $-0.2$
can be found in Fig.~\ref{betr}. Similar examples but with $\beta(r=0)\neq 0$ 
can be found in the appendix~C of \cite{Chae12}.
\begin{figure*} 
\begin{center}
\setlength{\unitlength}{1cm}
\begin{picture}(12,11)(0,0)
\put(-1.8,12.5){\includegraphics{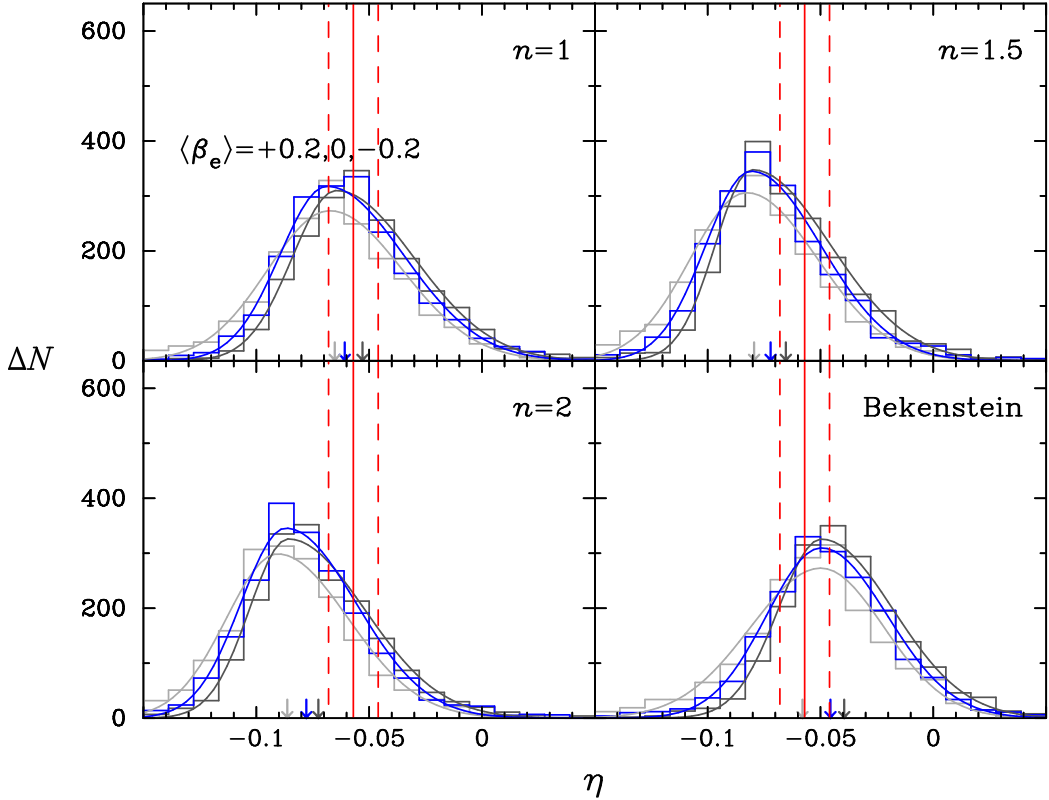}}
\end{picture}
\caption{
Same as Fig.~\ref{VDPconst} except that radially varying anisotropies are used 
and $\beta_{\rm e}$ refers to the radial average of an anisotropy within
$R_{\rm e}$ of a galaxy.
}
\label{VDPvar}
\end{center}
\end{figure*}

Fig.~\ref{VDPvar} shows the predicted distributions of $\eta$ for 
the same combinations of ($n$,~$\langle\beta_{\rm e}\rangle$) as in 
Fig.~\ref{VDPconst} of the constant anisotropy case. 
Compared with the constant anisotropy 
case the predicted mean $\langle\eta\rangle$ shifts by $\approx -0.01$ for
the same input of ($n$,~$\langle\beta\rangle$). Consequently,
the empirical value of $\eta$ is best matched by the case $n = 1$. 
The case $n = 2$ is now disfavoured. 
Interestingly, Bekenstein's model can now be consistent with the empirical 
value with a suitable choice of $\langle\beta_{\rm e}\rangle$ although 
it may not be a preferred model. These results confirm that anisotropy shapes
can matter in the study of MOND (as well as DM) in spheroidal galaxies.

The width (containing 68 percent) of the distribution ($\sim 0.27$) is 
about 20 percent larger compared with the constant anisotropy case, 
 but still somewhat lower than the measured standard deviation. 
However, the marginal discrepancy may be a result of non-Gaussian nature 
of the distribution.
Indeed, the standard deviation of the distribution ($\sim 0.4$) is larger than 
the 68 percent width and matches well the measured standard deviation.

\section{Discussion and Conclusions}

We have investigated VDPs for $r \la R_{\rm e}$ of $\sim 2000$ nearly round SDSS 
galaxies with $10 \la \log_{10}(M_\star/{\rm M}_\odot) \la 12$ 
(mean $\approx 11.3$).  
Previous study by \cite{CBK14} shows that the predicted distribution
of VDP slope $\eta$ (equation~\ref{eq:eta}) cannot match an observed 
distribution without DM under Newtonian gravity 
(see also Fig.~\ref{VDPDM}). Furthermore, as shown in 
\cite{CBK14} each galaxy requires different amount and profile of DM.
 DM fraction within $R_{\rm e}$ ranges from $f_{\rm DM}\approx 0.1$ -- $0.5$
(68\% range: see Fig.~\ref{Gdist}) and the inner DM density profile slope 
$\alpha$, in the generalised Navarro-Frenk-White (\citealt{NFW97}) 
parameterisation $\rho_{\rm DM}(r)\propto (r/r_s)^{-\alpha}(1+r/r_s)^{-3+\alpha}$, 
ranges from $\alpha\approx 0.7$ -- $1.5$ (68\%). 
These results mean that elliptical galaxies
do require DM under Newtonian gravity in a statistical sense, although some 
fraction of galaxies may not contain significant amount in the optical region.

Our analysis of the same galaxies under MOND here indicates that
the observationally-derived value of $\eta$ (in 11 nearly round 
ATLAS$^{\rm 3D}$ galaxies) can be reproduced based on a single MOND 
interpolating function without DM applied to all the galaxies.
It is particularly striking that with an interpolation function similar to
that of $n= 1$ or $2$ in equation~(\ref{eq:ifmun}), used in various MOND
studies particularly for disc galaxies (e.g.\ \citealt{FB05,SN07,Ang08,Mil12}), 
the empirical $\eta$ value can be easily matched with a relaxed
assumption of VD anisotropy (Figs.~\ref{VDPconst} and \ref{VDPvar}).

The specific functional form of MOND interpolating function depends, however, 
on the assumption on VD anisotropy (Figs.~\ref{VDPconst} and \ref{VDPvar}) and
cannot be uniquely determined from the current empirical $\eta$ value.
The implied interpolating function can be varied by varying mean anisotropy
or radial behaviours. For constant anisotropy and the 
functional form given by equation~(\ref{eq:ifmun}), mean anisotropy 
$\langle\beta\rangle=+0.2$, $0$ and $-0.2$ give, respectively, 
$0.7\la n\la 1.4$,  $1.1\la n\la 1.9$ and $1.3\la n\la 2.5$. For the
case of radially varying anisotropy there is an overall shift of 
$\Delta n \approx -0.5$ for the same value of mean anisotropy so that the 
traditionally standard model $n=2$ is disfavoured. These manifest a degeneracy 
between MOND interpolating function and VD anisotropy, which is reminiscent of 
the well-known mass-anisotropy degeneracy under standard dynamics 
(e.g.\ \citealt{BM82}). This interpolating function-anisotropy degeneracy 
can be alleviated by a more precise empirical value of $\langle\eta\rangle$ 
or prior constraints on anisotropy.

Our analysis has been limited to nearly round galaxies (under the spherical
 symmetry assumption) in the intermediate acceleration regime 
although they cover a broad range of mass and size. A MOND analysis of ETGs 
 of any ellipticity is considerably more challenging but would
be necessary for a more rigorous test of MOND in spheroidal systems. 
If a random sample of elliptical galaxies (drawn from the same SDSS 
parent sample of ETGs using the same criteria except for the ellipticity limit)
 is analysed based on the MOND spherical Jeans 
equation (equation~\ref{eq:jeans}), the resulting $\eta$-distribution has 
a similar mean but a somewhat larger scatter, probably hinting that MOND 
works also for ETGs of any ellipticity.  
 
\begin{figure} 
\begin{center}
\setlength{\unitlength}{1cm}
\begin{picture}(9,7)(0,0)
\put(-0.7,7.4){\includegraphics{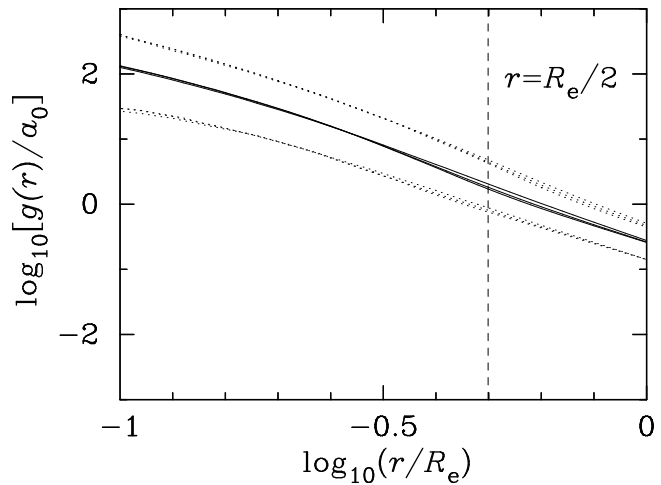}}
\end{picture}
\caption{
Distribution of gravitational acceleration $g(r)$ normalised by 
$a_0=1.2\times 10^{-10}{\rm m}~{\rm s}^{-2}$ due to stellar mass within radius 
$r$ in $\sim$ 2000 nearly round SDSS galaxies based on modelling results with
$n=1$, $1.5$ and $2$ for $\langle\beta\rangle=0$ shown in Fig.~\ref{VDPconst}.
The solid curves represent the medians while the dotted curves represent the 
95\% limits.
The observed VDs of SDSS galaxies analysed in this work are luminosity-weighted
 LOSVDs within projected radius $R_{\rm ap}$, which is peaked at $R_{\rm e}/2$
(represented by the vertical dashed line) 
but ranges from $\sim 0.1R_{\rm e}$ up to $\sim R_{\rm e}$.
}
\label{grange}
\end{center}
\end{figure}

Our analysis provides a novel and rigorous test of MOND in spheroidal galaxies 
in the sense that it is based on a statistically representative sample of 
spheroidal galaxies and a range of possibilities of VD anisotropy. 
Our test is most sensitive at $r\la R_{\rm e}/2$ where gravitational 
acceleration $g$ due to the baryonic mass distribution is 
 $\ga a_0$ (Fig.~\ref{grange}). However, our galaxies cover the range
$0.1R_{\rm e}\la r \la R_{\rm e}$ (see Fig.~1 of \citealt{CBK14}) for which
$0.3a_0 \la g \la 100 a_0$ (Fig.~\ref{grange}). 
Our results support the view that MOND suggested by dynamics of rotating 
galaxies is also likely to be valid for  dispersion-supported 
galaxies in their inner regions ($r \la R_{\rm e}$).
 Our results are in line with the test of MOND through 
hydrostatic equilibrium in two X-ray bright elliptical galaxies 
over a wide acceleration range by \cite{Mil12}. 

In this work we could not separately consider slow and fast rotators because
the necessary kinematic information is not available for the analysed SDSS 
galaxies. The fact that the similarly selected ATLAS$^{\rm 3D}$ elliptical 
galaxies are nearly evenly divided (see section~2) hints that the SDSS sample 
may also contain both kinematic classes, although it is likely 
to be somewhat biased towards slow rotators as the galaxies are on average more 
massive and larger than the ATLAS$^{\rm 3D}$ counterparts (see Fig.~\ref{Gdist}).
Recently, \cite{Sam14} carried out individual Jeans modelling of ten elliptical 
galaxies, four of whom are slow rotators, based on observed GC kinematic data
up to several effective radii. Considering two cases of 
constant anisotropy $\beta=0$, $-0.2$ and one case of radially varying  
$\beta(r)\approx 0.5 r/(r+1.4 R_{\rm e})$ for MOND interpolating functions of 
$n=1$, $2$ in equation~(\ref{eq:ifmun}) and that of equation~(\ref{eq:ifmubek}),
 \cite{Sam14} could not find successful fits of slow rotators without 
additional component of DM beyond $\sim 2$-$3 R_{\rm e}$. For one galaxy
 NGC 4486 (M87), which is in the central region of a cluster, 
the considered MOND models and anisotropies had difficulty of
fitting GC kinematic data even in the inner region ($r\la 0.5 R_{\rm e}$).
 However, as \cite{Sam14} notes, it remains unclear whether these problems
based on limited cases of anisotropies imply breakdown of MOND for certain
objects. In this work we have considered a broader range of possibilities of
anisotropy and found that VDPs of a mixed population of fast and slow rotators 
for $r\la R_{\rm e}$ could be statistically explained by MOND without any DM. 

If the empirical MOND law is to be truly meaningful, there must exist a 
universal interpolating function. Studies up to the present indicate that 
this is a reasonable possibility. 
However, a firm conclusion can only be reached through precise determination of
 interpolating functions of galaxies of various types. Our analysis shows
that the statistics of VDPs of elliptical galaxies provides a useful tool to
test interpolating functions. This statistical analysis can complement 
individual analyses that can be performed with detailed kinematic data 
(e.g.\ \citealt{Sam14}). 
Despite its current limits, our analysis appears to support the simple 
interpolating function [equation~(\ref{eq:ifmun}) with $n=1$]
(see Figs.~\ref{VDPconst} and \ref{VDPvar}), which has also 
been preferred by galaxy rotation data (e.g.\ \citealt{FB05,SN07}) and 
 studies of elliptical galaxies [see \cite{Mil12} and references therein],  
although we cannot yet rule out the traditionally standard model with $n=2$ or
even the interpolating function (equation~\ref{eq:ifmubek})
implied by Bekenstein's modified gravity (\citealt{Bek04,ZF06}).
 
If a single interpolating function turns out to explain dynamics of all 
galaxies, that would have far-reaching implications. In this respect it would
be quite interesting to empirically verify and determine the supposed
universal MOND interpolating function with future data. 
For spheroidal galaxies both statistical (for large samples, as was 
demonstrated in this study) and individual (for large radial extent, performed 
earlier and ongoing) analyses of VDPs will be useful.

\bigskip

We are grateful to Srdjan~Samurovi\'{c} for providing a thorough and 
helpful report of the submitted manuscript. We also thank Maurice van 
Putten for useful discussions/comments and Mariangela Bernardi
for helpful communication regarding SDSS galaxies.

\bibliographystyle{mn2e}

\end{document}